\documentstyle [preprint,revtex,eqsecnum]{aps}

\global\arraycolsep=2pt			

\makeatletter
\def\footnotesize{\@setsize\footnotesize{9.5pt}\xpt\@xpt
\abovedisplayskip 10pt plus2pt minus 5pt
\belowdisplayskip \abovedisplayskip
\abovedisplayshortskip \z@ plus 3pt
\belowdisplayshortskip 6pt plus 2pt minus 2pt
\def\@listi{\topsep 6pt plus 2pt minus 2pt
\parsep 3pt plus 2pt minus 1pt \itemsep \parsep}}
\makeatother

\begin {document}

\preprint {UW/PT-92-16}

\begin {title}
    {%
	Summing tree graphs at threshold
    }%
\end {title}

\author {Lowell S. Brown}

\begin {instit}
    {%
    Department of Physics, FM-15,
    University of Washington,
    Seattle, Washington 98195
    }%
\end {instit}

\begin {abstract}
    {%
	The solution of the classical field equation generates the
	sum of all tree graphs. We show that the classical equation
	reduces to an easily solved ordinary differential equation
	for certain multiparticle threshold amplitudes and compute
	these amplitudes.
    }%
\end {abstract}

\narrowtext

\newpage

    \ifpreprinsty
    \vbox to \vsize
	{%
	\vfill
	\baselineskip .28cm
	\par
	\font\eightrm = cmr8
	\eightrm \noindent
	This report was prepared as an account of work sponsored by the
	United States Government.
	Neither the United States nor the United States Department of Energy,
	nor any of their employees, nor any of their contractors,
	subcontractors, or their employees, makes any warranty,
	express or implied, or assumes any legal liability or
	responsibility for the product or process disclosed,
	or represents that its use would not infringe privately-owned rights.
	By acceptance of this article, the publisher and/or recipient
	acknowledges the U.S. Government's right to retain a non-exclusive,
	royalty-free license in and to any copyright covering this paper.%
	}%

\newpage

The important and outstanding problem of high-energy baryon number
violation \cite{mattis} motivates the investigation of simplier
processes which share some of its features. One feature of the
high-energy baryon violation is the production of a very large number
of particles. Even in a weakly-coupled theory, many-particle
amplitudes may become large because they involve a large number of
graphs. Cornwall \cite{cornwall} and Goldberg \cite{goldberg} have
examined the lowest-order, tree graph amplitudes for many particle
production. Recently, Voloshin \cite{voloshin} has considered the tree
graphs in the simple, unbroken $\lambda \, \phi^4$ theory for the
amplitude where a highly off-mass-shell $\phi$-field produces a large
number $n$ of on-mass-shell $\phi$-particles. In particular, he
considered the threshold limit in which all the produced particles are
at rest and obtained an exact result for this amplitude by deriving
and solving a recursion relation for the many-particle amplitudes.
Using this technique, Argyres, Kleiss, and Papadopoulos extended
Voloshin's result to include the case of the spontaneously broken
$\lambda \, \phi^4$ theory which contains cubic as well as quartic
interactions. The purpose of the present note is to point out that
these previous results are obtained very simply if one recalls that
the generating function for the tree graphs is the solution to the
classical field equation \cite{me}. We shall also show how to
generalize the results to the case of the unbroken, multicomponent
$O(N)$ scalar theory.

Our object is to compute the amplitudes for the field $\phi$ to create
$n$ particles out of the vacuum,
$\langle n | \phi | 0 \rangle$, for the simple scalar theory described
by the Lagrange function
\begin{equation}
{\cal L} = - {1\over2} (\partial \phi)^2 -{1\over2} m^2 \phi^2 -
{1\over4!} \lambda \phi^4 + \rho \phi \,.
\end{equation}
The source $\rho$ is introduced so as to generate these amplitudes
according to the usual reduction formula method \cite{brown},
\begin{equation}
\langle n | \phi(x) | 0 \rangle = \prod_{a=1}^n \, \int (d^4x_a) \,
e^{-i p_a x_a} \, (p_a^2 + m^2) \,\left.  {\delta \over \delta \rho(x_a)}
\langle 0 {+}  | \phi(x) |  0 {-} \rangle^{\rho} \, \right|_{\rho =0} \,.
\label{fun}
\end{equation}
The tree graph approximation is obtained by the replacement
\begin{equation}
\langle 0 {+} | \phi(x) | 0 {-} \rangle^{\rho} \to \phi_{\rm cl}(x) \,,
\end{equation}
where $\phi_{\rm cl}$ is the solution to the classical field equation
driven by the source $\rho$,
\begin{equation}
(-\partial^2 + m^2) \phi_{\rm cl} + {1\over3!} \lambda \phi_{\rm cl}^3
= \rho \,,
\label{cleq}
\end{equation}
subject to the quantum time-ordered boundary conditions which give the
prescription $m^2 \to m^2 - i \epsilon$ in the propagator. This defines
the classical field $\phi_{\rm cl}$ as a functional of the source,
$\phi_{\rm cl} =  \phi_{\rm cl}[\rho]$.
We shall consider only the threshold limit, ${\bf p}_a = 0$. In this
limit, the space-time dependent source $\rho(x)$ and the resulting
field $\phi_{\rm cl}(x)$ may be replaced by
spatially uniform but time-dependent functions $\rho(t)$ and
$\phi_{\rm cl}(t)$.  Thus the field equation (\ref{cleq}) reduces to an
ordinary
differential equation. The mass-shell amplitude is obtained by setting
\begin{equation}
\rho(t) = \rho_0 \, e^{i \omega t}
\end{equation}
and then taking the limit $\omega \to m$. To see how this goes, we
first examine the solution to the classical field equation
({\ref{cleq}) when there is no interaction ($\lambda \to 0$),
\begin{equation}
\phi_{\rm cl} \to z(t) \equiv z_0
	\, e^{i \omega t} \,,
\label{lim}
\end{equation}
where
\begin{equation}
z_0 = {\rho_0 \over m^2 - i \epsilon - \omega^2} \,.
\label{cons}
\end{equation}
We now insert the expansion of $\phi_{\rm cl}$ in powers of the
coupling $\lambda$
\begin{equation}
\phi_{\rm cl} = z + \lambda \, \phi_{\rm cl}^{(1)} +
\lambda^2 \, \phi_{\rm cl}^{(2)} + \cdots
\end{equation}
into the field equation (\ref{cleq}) and identify the coefficients of
the various powers of the coupling $\lambda$. This shows that
$\phi_{\rm cl}^{(1)}$ is proportional to $\lambda \, z^3 /3!$.
Similarly, proceeding to higher orders in the
expansion shows that $\phi_{\rm cl}$ is an ordinary function of $z(t)$,
$\phi_{\rm cl}(t) = \phi_{\rm cl}(z(t))$.
Hence, in view of Eqs.~(\ref{lim}) and (\ref{cons}), the functional
derivatives which
occur in the reduction formula (\ref{fun}) become ordinary derivatives:
\begin{equation}
\int (d^4x_a) \, e^{i \omega t_a} \, (m^2 - \omega^2)
\, {\delta \over \delta \rho(x_a)} \phi_{\rm cl}(t;\, [\rho])
= {\partial \over \partial z_0} \phi_{\rm cl}(z(t)) \,.
\end{equation}
We see that the threshold $n$-particle amplitude in
the tree approximation may be expressed as
\begin{equation}
\langle n | \phi(0) | 0 \rangle^{\rm tree}_{\rm threshold}
= \left. \left( {\partial
\over \partial z} \right)^n \, \phi_{\rm cl} \, \right|_{z=0} \,.
\label{result}
\end{equation}
To go on mass shell, $\omega \to m$, we take $\rho_0 \to 0$ in such a
way as to keep $z(t)$ finite. In this limit, the classical field obeys the
homogeneous, ordinary differential equation
\begin{equation}
\left[ {d^2 \over dt^2} + m^2 \right] \phi_{\rm cl}(t) + {1\over3!}
\lambda \phi_{\rm cl}^3(t) = 0 \,,
\label{diffeq}
\end{equation}
with the condition that $\phi_{\rm cl}(t)$ approaches $z(t)$ as $\lambda$
vanishes.

\subsection{Unbroken Symmetry}

The ordinary differential equation (\ref{diffeq}) has a constant
of the motion, the energy integral
\begin{equation}
{1\over2} \left( {d \phi_{\rm cl} \over dt} \right)^2
+ {1\over2} m^2 \, \phi_{\rm cl}^2 + {1\over4!} \lambda \,
\phi_{\rm cl}^4 = E^2 \,.
\end{equation}
Since $\phi_{\rm cl}$ contains only an ascending power series in the
oscillating function $z(t)$, the left-hand side of this equation
contains only oscillating terms. Hence the constant $E$ must vanish, and
the final integration to obtain $\phi_{\rm cl}$ gives a simple elementary
function. Rather than indicating the intermediate steps, we shall just
write down the result since it is easier to verify directly that it is
the proper solution of the original differential equation (\ref{diffeq}):
\begin{equation}
\phi_{\rm cl}(t) = {z(t) \over 1 - (\lambda / 48 \, m^2) z(t)^2 } \,.
\label{sol}
\end{equation}
This is the function for the case of unbroken symmetry with $m^2 > 0$.
Placing the solution (\ref{sol}) in Eq.~(\ref{result}) shows that the
amplitude vanishes unless $n = 2k + 1$ is odd, where
\begin{equation}
\langle 2k + 1 | \phi(0) | 0 \rangle^{\rm tree}_{\rm threshold} =
(2k + 1)! \left( {\lambda \over 48 \, m^2} \right)^k \,.
\label{unbreak}
\end{equation}
This is the result of Voloshin \cite{voloshin}.

\subsection{Broken Symmetry}

The reflection symmetry $\phi \to - \phi$ is broken when
$m^2 \to -m^2 < 0$. In
this case the field equation (\ref{diffeq}) has the constant solution
\begin{equation}
\phi_{\rm cl} \to \phi_0 = \sqrt{ {3! \, m^2 \over \lambda} } \,.
\end{equation}
Expanding the field about this constant solution gives rise to 3-field
as well as 4-field couplings, and the graphical structure becomes more
complex. Moreover, the mass parameter in the Lagrange function for the
shifted field is altered to $m_1 = \sqrt{2} \, m$ and so we must now
expand about
\begin{equation}
z(t) = z_0 \, e^{i m_1 t} \,.
\end{equation}
Again since an energy integral to the equation of
motion exists, the classical field equation (\ref{diffeq}) may be solved.
Rather than writing down the intermediate steps, it again suffices to
display the solution since its verification is simple:
\begin{equation}
\phi_{\rm cl}(t) = \phi_0 {1 + z(t) / 2\phi_0 \over 1 - z(t) / 2\phi_0} \,.
\end{equation}
Inserting this solution into Eq.~(\ref{result}) gives
\begin{equation}
\langle n | \phi(0) | 0 \rangle^{\rm tree}_{\rm threshold} =
n! \left( {1 \over 2\phi_0} \right)^{n-1} \,.
\end{equation}
This is the result of Argyres et al. \cite{argyres}.

\subsection{$O(N)$ Model}

The results of the unbroken theory are easily extended to the $O(N)$
theory in which the single $\phi$ field is replaced by a
vector field $\phi^a$ with $N$ components, and the interaction now
involves the $O(N)$ invariant $(\phi^a \phi^a)^2$. It is
straightforward to check that the previous solution (\ref{sol})
generalizes to
\begin{equation}
\phi_{\rm cl}^a(t) = {z^a(t) \over 1 - (\lambda / 48 \, m^2)
	 z(t) \cdot z(t) } \,.
\end{equation}
In the previous simple one-component theory, the squared matrix
element, divided by the Bose symmetrization factor $n!$ and multiplied
by the appropriate phase space factor, gives the threshold limit of the
absorptive part of the $\phi$-field propagator. The analogous
construction for the
$O(N)$ theory involves some combinatorial analysis which, as we now
show, is simplified by using functional methods. For the $O(N)$
theory, the absorptive part of the propagator in the tree
approximation entails the schematic structure
\begin{equation}
M^{ab}(2k+1) = \left. \sum \langle 0 | \phi^a | 2k+1 \rangle
\langle 2k+1 | \phi^b | 0 \rangle \right|^{\rm tree}_{\rm threshold} \,.
\label{sum}
\end{equation}
Here the sum is over all states with the total particle number $n =
2k+1$ including implicit Bose symmetrization factors.
To perform this sum with the correct factors, we note that the
exponential operation applied to
a squared matrix element in our $z$ representation,
\begin{equation}
\exp\left\{{\partial\over\partial z^{*a}}{\partial\over\partial z^a}\right\}
= \sum_{l=0}^\infty {1 \over l!}
\left( {\partial\over\partial z^{*a}} \right)^l
\left( {\partial\over\partial z^a} \right)^l \,,
\end{equation}
(with no sum over $a$) produces a sum over all particle numbers $l$ of
particle type $a$ in the resulting intermediate state with the correct
Bose symmetrization factor of $l!$ in the denominator. Hence the operation
\begin{equation}
\exp\left\{{\partial\over\partial z^{*1}}{\partial\over\partial z^1}\right\}
\cdots
\exp\left\{{\partial\over\partial z^{*N}}{\partial\over\partial z^N}\right\}
=
\exp\left\{{\partial\over\partial  z^*} \cdot {\partial\over\partial
 z}\right\}
\end{equation}
applied to monomial
$$
\left({\lambda \over 48 \, m^2} \right)^{2k} z^{*a} ( z^* \cdot
 z^*)^k  z^b ( z \cdot  z)^k \,,
$$
which is the order $2k+1$ term in the expansion of $\phi_{\rm cl}^{*a} \,
\phi_{\rm cl}^b$, produces the sum (\ref{sum}) in the limit
$z^{*a} = z^b =0$. Since
\begin{equation}
\exp\left\{ y {\partial \over \partial x} \right\} f(x) = f(x+y) \,,
\end{equation}
we have
\begin{equation}
M^{ab}(2k+1) = \left({\lambda \over 48 \, m^2} \right)^{2k} {\partial
\over \partial z^a} \left( {\partial \over \partial  z} \cdot
{\partial \over \partial  z} \right)^k
	z^b ( z \cdot  z)^k \,.
\end{equation}
The derivatives which appear here may be evaluated recursively by using
\begin{equation}
\left( {\partial \over \partial  z} \cdot
{\partial \over \partial  z} \right) z^a ( z \cdot  z)^l
= 4l \, ( l + N/2) \, z^a ( z \cdot  z)^{l-1} \,.
\end{equation}
Thus
\begin{equation}
M^{ab}(2k+1) = \delta^{ab} \left({\lambda \over 48 \, m^2}\right)^{2k} 4^k k!
	\, {\Gamma(k + 1 + N/2) \over \Gamma(1+N/2) } \,.
\end{equation}

\subsection{Quantum Mechanics}

The validity of the tree approximation is illustrated by considering
the quantum mechanical analog of the field theory --- the field theory
in zero spatial dimensions --- the anharmonic oscillator. In this
case, the ``threshold limit'' gives the full amplitude. Dividing the
previous result for the unbroken theory (\ref{unbreak}) by
$\sqrt{(2k+1)!(2m)^{2k+1}}$ to produce the normalized quantum mechanical
amplitude and then squaring to obtain the spectral weight gives
\begin{equation}
r_{2k+1}^{\rm tree} = \left| \langle 2k + 1 | q | 0 \rangle^{\rm tree}
\right|^2 =
{1\over 2m} \, \left({\lambda \over 96 \, m^3}\right)^{2k} \, (2k+1)! \,.
\end{equation}
Although the coupling $\lambda$ may be small, this amplitude becomes
large for a sufficiently highly excited state where $k$ is very large,
and the tree approximation must break down. Indeed, inserting the
complete set of intermediate energy eigenstates into the ground state
commutator matrix element
\begin{equation}
\langle 0 | [q,\, dq/dt] | 0 \rangle = i
\end{equation}
yields the sum rule
\begin{equation}
\sum\nolimits_n 2E_n \, r_n = 1 \,,
\end{equation}
where we have chosen the energy scale to make the ground state energy
vanish. Since a sum of positive terms appears here, and since the
anharmonic coupling increases the energy of an intermediate state, we
have the bound\footnote{
The inequality (\ref{ineq}) can obviously be
sharpened by placing a factor of $(1 - 2E_1 \, r_1)$ on the right-hand
side. In the weak coupling limit, this factor is of order
$(\lambda / m^3)^2$. This, however, does not alter the leading
behavior of the large $k$ restriction shown in Eq.~(\ref{last}).}
\begin{equation}
r_n < {1 \over 2E_n} < {1 \over 2m} {1 \over n} \,.
\label{ineq}
\end{equation}
Hence even for weak coupling, the tree approximation must break down
for highly excited states where
\begin{equation}
(2k+1) (2k+1)! \sim \left( {96 \, m^3 \over \lambda} \right)^{2k} \,,
\end{equation}
or, on using Stirling's approximation for the factorial, when
\begin{equation}
k \sim (48 \, m^3 /\lambda) \, e \,,
\label{last}
\end{equation}
where $e$ is the base of the natural logarithm. This restriction just
states that the tree approximation must break down for states which
are so highly excited that the anharmonic interaction becomes
comparable to the harmonic term, $m^2 \, \langle q^2 \rangle \sim \lambda \,
\langle q^4 \rangle $ since, in the harmonic approximation, $ \langle
q^2 \rangle \sim k/m$ and $\langle q^4 \rangle \sim k^2 / m^2$.

\newpage

\null
\bigskip\bigskip

I have enjoyed conversations on this topic with D. G. Boulware,
H. Goldberg, L. D. McLerran, and L. G. Yaffe. This work was supported,
in part, by the U. S. Department of Energy under grant DE-AS06-88ER40423,
and it was completed at the Aspen Center for Physics.


\begin {references}

\bibitem{mattis}
	A recent review appears in
	M. P. Mattis,
	Phys. Rep. {\bf 214}, 159 (1992).

\bibitem{cornwall}
	J. M. Cornwall,
	Phys. Lett. B {\bf 243}, 271 (1990).

\bibitem{goldberg}
	H. Goldberg,
	Phys. Lett. B {\bf 246}, 445 (1990);
	Phys. Rev. D {\bf 45}, 2945 (1992).

\bibitem{voloshin}
	M. B. Voloshin,
	Minnesota preprint TPI-MINN-92/1-T,
	Nucl. Phys. B, to be published.

\bibitem{argyres}
	E. N. Argyres, R. H. P. Kleiss, and C. G. Papadopoulos,
	CERN preprint CERN-TH.6496.

\bibitem{me}
	Y. Nambu,
	Phys. Lett. B {\bf 26}, 626 (1968);
	D. G. Boulware and L. S. Brown,
	Phys. Rev. {\bf 172}, 1628 (1968).

\bibitem{brown}
	The reduction formula and the relation of tree graphs to the
	classical field is discussed, for example, in
	L. S. Brown,
	{\it Quantum Field Theory},
	Cambridge University Press, 1992.

\end {references}
\end {document}